\documentclass[conference]{IEEEtran}
\IEEEoverridecommandlockouts
\usepackage{cite}
\usepackage{amsmath,amssymb,amsfonts}
\usepackage{algorithmic}
\usepackage{graphicx}
\usepackage{textcomp}
\usepackage{xcolor}
\def\BibTeX{{\rm B\kern-.05em{\sc i\kern-.025em b}\kern-.08em
    T\kern-.1667em\lower.7ex\hbox{E}\kern-.125emX}}

\begin{document}

\title{Digital Twins for Moving Target Defense Validation in AC Microgrids}

% \author{
% \IEEEauthorblockN{\textbf{Suman Rath}\IEEEauthorrefmark{1}, \textbf{Subham Sahoo}\IEEEauthorrefmark{2}, \textbf{Shamik Sengupta}\IEEEauthorrefmark{1}}
% \IEEEauthorblockA{
% \IEEEauthorblockA{\IEEEauthorrefmark{1}Department of Computer Science and Engineering, University of Nevada, Reno, USA\\
% \IEEEauthorrefmark{2}Department of Energy, Aalborg University, Denmark\\
% }
% E-mail: \{srath@nevada., ssengupta@\}unr.edu, sssa@energy.aau.dk}
% }

\author{\IEEEauthorblockN{Suman Rath}
\IEEEauthorblockA{\textit{Department of Computer Science}\\ \textit{and Engineering} \\
\textit{University of Nevada, Reno}\\
Reno, NV 89557, USA \\
E-mail: srath@nevada.unr.edu}
\and
\IEEEauthorblockN{Subham Sahoo}
\IEEEauthorblockA{\textit{Department of Energy}\\ \textit{(AAU Energy)} \\
\textit{Aalborg University}\\
Aalborg, 9220, Denmark \\
E-mail: sssa@energy.aau.dk}
\and
\IEEEauthorblockN{Shamik Sengupta}
\IEEEauthorblockA{\textit{Department of Computer Science}\\ \textit{and Engineering} \\
\textit{University of Nevada, Reno}\\
Reno, NV 89557, USA \\
E-mail: ssengupta@unr.edu}}

\maketitle

\begin{abstract}
Cyber-physical microgrids are vulnerable to stealth attacks that can degrade their stability and operability by performing low-magnitude manipulations in a coordinated manner. This paper formulates the interactions between CSAs and microgrid defenders as a non-cooperative, zero-sum game. Additionally, it presents a hybrid Moving Target Defense (MTD) strategy for distributed microgrids that can dynamically alter local control gains to achieve resiliency against Coordinated Stealth Attacks (CSAs). The proposed strategy reduces the success probability of attack(s) by making system dynamics less predictable. The framework also identifies and removes malicious injections by modifying secondary control weights assigned to them. The manipulated signals are reconstructed using an Artificial Neural Network (ANN)-based Digital Twin (DT) to preserve stability. To guarantee additional immunity against instability arising from gain alterations, MTD decisions are also validated (via utility and best response computations) using the DT before actual implementation. The DT is also used to find the minimum perturbation that defenders must achieve to invalidate an attacker's knowledge effectively.
\end{abstract}


\section{Introduction}
Microgrids are the gateways to carbon neutrality in the power and energy sector. CSAs are often depicted as threats to the stability and control framework of cyber-physical microgrids \cite{9777754}. The main factor contributing to the undetectability of CSAs is their ability to analyze system dynamics and leverage the same to achieve persistence against generic defenses (e.g., bad data detectors) \cite{rath2022behind}.
MTD can enhance the security of microgrids against such attacks \cite{liu2021converter} via deceptive control by continually changing its configuration to introduce an unpredictability element in system dynamics that remains invisible from an adversarial perspective
\cite{9777754}.
The continuous alteration of the system configuration also makes it difficult for malicious agents to exploit inherent vulnerabilities and hide potential manipulations, thereby increasing their detection probability and alleviating the possibility of damage. Thus, MTD formulations can allow such systems to continue operating normally in adversarial settings. However, the implementation of such dynamic response frameworks often warrants a careful analysis of the trade-off between security and operational performance \cite{9475527}.

Miscalculated defender actions can have a counter-intuitive effect on microgrid stability and degrade its performance by contributing to the attacker's intentions. To avoid such consequences, MTD implementations must be done in a strategic manner through detailed cost-benefit evaluations before the execution of defender decisions in real-time \cite{9144465}.
This paper presents a DT-enabled MTD strategy that can periodically alter primary control gains to reduce the success probability of  CSAs. The proposed strategy also includes an event-triggered mitigation mechanism that can actively respond to successful attack vectors (if any) by altering communication weights to reject malicious signals. The presented framework uses an ANN-based DT to reconstruct the manipulated signal to preserve system stability and achieve complete resiliency. The DT is also used to analyze cost-benefit trade-offs and predict the system's response to MTD decisions (before actual implementation) in an intelligent manner. We present several simulation results to demonstrate the working principle of CSAs against generic bad data detectors and analyze the performance of the proposed strategy against the same.

\section{Background Knowledge}

\subsection{Cyber-Physical Vulnerabilities in Microgrids}
Microgrids are typically controlled by a hierarchical framework consisting of primary, secondary, and tertiary layers. The primary and secondary layers are responsible for nodal synchronization, and voltage/frequency regulations. The secondary controller relies on a communication-dependent computing setup to achieve its control goals \cite{rath2020cyber}. This setup exposes the microgrid control framework to cyber vulnerabilities through which attack vectors can maliciously manipulate the system state trajectory. Apart from cyber vulnerabilities, microgrids are also vulnerable to physical attack vectors that seek to manipulate sensors, embedded devices, and/or actuators to achieve measurement-level alterations. Physical devices with embedded malware may be used to send erroneous measurement signals into the control plane forcing it to make wrong decisions. This creates a series of cascading failures threatening nominal microgrid operations \cite{rath2022behind}.

Additional vulnerabilities in microgrids can include weak communication protocols, expired security certificates, and code loopholes \cite{rath2022behind}. Weak communication protocols can lead to packet losses and allow attackers to decode state information. Expired security certificates can allow attackers to masquerade abnormal attack-type data traffic as normal phenomena. Code loopholes can allow attackers to snoop on state information via hidden back doors by exploiting the defender's inability to detect malicious data traffic. Adversaries generally use several vulnerabilities discussed in this subsection to port malware (e.g., rootkits \cite{rath2022behind}) into the distributed generators (DGs) in the microgrid network. These malware variants can introduce stealthy attack vectors like CSAs into the system. More details about the characteristics and working mechanism of microgrid CSAs are provided in the following subsection.

\subsection{CSAs: Characteristics and Operation}
CSAs are attack vectors that can be executed in the microgrid environment by kernel-level malware like rootkits \cite{rath2022behind}. These attack vectors introduce manipulations from one or more node(s) in the network and attempt to hide them by introducing complementary injections from the other nodes. Further, the magnitude of each false data injection is kept bounded so as to evade bad data detectors. Low-magnitude manipulations can also be masqueraded as noise or other natural disturbances. This allows the CSA to hide its actions from microgrid operators and defenders. The operating principle of CSAs varies as per the locations at which the executing malware is lodged.

CSAs can modify measurement-level signal values if the executing malware has access to sensor-level devices. These manipulations can then propagate through the communication plane into the control layer to force the system into generating erroneous control signals. This can lead to serious repercussions (e.g., frequency/voltage instability, blackouts, etc.), especially in autonomous microgrids which don't have access to grid-mandated parameter setpoints. In autonomous microgrids, parameter setpoints are typically enforced by the secondary control framework. If the malware is lodged at the control layer, such setpoints can also be altered to serve the attacker's intentions \cite{rath2022behind}. The CSA modeling approach adopted in this paper is described in the following section.

\section{Control Structure and Attack Formulation}
The AC microgrid setup utilized in this paper has a distributed architecture and is regulated using a hierarchical two-layered control framework. A detailed description of the microgrid control framework is provided below.

\begin{figure}
\centering
\centerline{{\includegraphics[width=\linewidth]{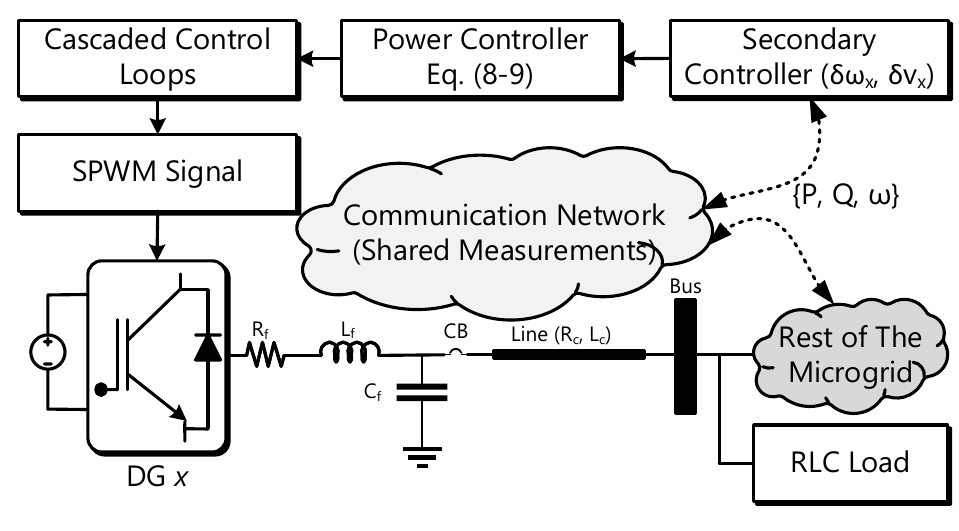}}}
\caption{{Microgrid control architecture.}}
\label{fig:two}
\end{figure}

\subsubsection{Primary Control}
The primary control layer strives to achieve voltage and frequency synchronization among the microgrid nodes so as to regulate active and reactive power sharing in the microgrid network. To achieve its objectives, the controller utilizes a droop-based action which is depicted by the following primary power control equations:
\begin{equation}
\omega^{\ast}_x = \omega_{nom}-D_{P_x}P_x
\end{equation}
\begin{equation}
v^{\ast}_{x} = v_{nom}-D_{Q_x}Q_x
\end{equation}
where \(\omega^{\ast}_x\) represents the magnitude of local frequency at the $x^{th}$ DG and \(v^{\ast}_{x}\) represents voltage magnitude, \(\omega_{nom}\) and \(v_{nom}\) represent nominal frequency and nominal voltage respectively, \(P_x\) represents active power measurement and \(Q_x\) represents reactive power measurement. \(D_{P_x}\) and \(D_{Q_x}\) are droop coefficients which are formulated as:
\begin{equation}
    D_{P_1}P_1 = D_{P_2}P_2 = ... = D_{P_N}P_N = \Delta\omega_{T}
\end{equation}
\begin{equation}
    D_{Q_1}Q_1 = D_{Q_2}Q_2 = ... = D_{Q_N}Q_N = \Delta{v}_{T}
\end{equation}
where \(\Delta v_{T}\) and \(\Delta{\omega}_{T}\) are the maximum allowable deviations for voltage and frequency signals.
The droop-based action of the local DG-level primary controller causes the frequency and system voltage to drop below the nominal magnitude as $D_P{_{\circ}}P_{\circ}$ and $D_Q{_{\circ}}Q_{\circ}$ values increase. To eliminate the impact of this droop, a secondary control layer is used that utilizes a cooperative synchronization-based framework to remove the droop and restore deviated signal(s) to their nominal path(s) while keeping power-sharing regulations intact.

\subsubsection{Secondary Control}
The secondary control layer considered in this paper uses a cooperative, leader-follower synchronization strategy where one DG is chosen to hold the reference parameter values, and all the other DGs have to track this reference DG to remove the droop in their local frequency and voltage values caused due to the primary controller's action. The target objectives for the secondary control layer are depicted below:
\begin{equation}
\lim_{t \to \infty}||\omega_x(t)-\omega_n|| = 0 \;\forall\; x
\end{equation}
\begin{equation}
\lim_{t \to \infty}||D_{P_x}{P_x}-D_{P_y}{P_y}|| = 0 \;\forall\; x,\;y
\end{equation}
\begin{equation}
\lim_{t \to \infty}||D_{Q_x}{Q_x}-D_{Q_y}{Q_y}|| = 0 \;\forall\; x,\;y
\end{equation}
To achieve the above objectives, the secondary layer employs a set of distributed local controllers, each of which is associated with a DG and receives shared measurements from other DGs in the network. As shown in Fig. \ref{fig:two}, after the receipt of neighboring measurements, the local controller computes a set of two error signals $\{\delta \omega, \delta v\}$ which are fed to the primary power control equations depicted in (1) and (2). The updated power control equations are depicted below:
\begin{equation}
\omega^{\ast}_x = \omega_{nom}-D_{P_x}P_x+\delta \omega_x
\end{equation}
\begin{equation}
v^{\ast}_{x} = v_{nom}-D_{Q_x}Q_x+\delta v_x
\end{equation}
where $\delta \omega_x$ and $\delta v_x$ are computed using:
\[
\delta\dot{\omega_x} = K_1\Big(\sum_{y\epsilon{N(x)}}{a_{xy}}(\omega_{y}-\omega_x)+g_x(\omega_n-\omega_x)+
\]
\begin{equation}
\sum_{y\epsilon{N(x)}}{a_{xy}}(D_{P_y}P_{y}-D_{P_x}P_{x})\Big)
\end{equation}
\begin{equation}
\delta\dot{v_x} = K_2\Big(\sum_{y\epsilon{N(x)}}{a_{xy}}(D_{Q_y}Q_{y}-D_{Q_x}Q_{x})\Big)
\end{equation}
where $a_{xy} \in A$ is an element of the adjacency matrix $A$ that represents a bidirectional communication graph connecting DGs in the network.
%$a_{xy}$ signifies the weight assigned to the communication link connecting DGs $x$ and $y$.
$g$ is the pinning gain and \(K_1, K_2\) are parameters of constant magnitude.
In this framework, attackers generally attempt to introduce manipulations into the measurement and communication layers. The manipulations are often aimed at disrupting the nominal state trajectory.

\subsubsection{Modeling of CSAs}
Microgrid attackers generally attempt to infect the maximum possible number of nodes in the network so as to leverage them for modifying the nominal state trajectory. One of the attack variants that attackers can attempt to execute is a CSA where simultaneous injections are introduced into the network from multiple DGs. These injections are often introduced in a coordinated strategy so as to trick the defender into having a misconception that the microgrid is in its nominal state. In the event of a CSA, the microgrid state is forced to validate an attack signal into the generic control framework. The altered control vector $c_a$ can be modeled as:
\begin{eqnarray}
{{c_a}} = f({x_{nom}}) +  {B} {x}_{a}
\end{eqnarray} 
where ${x_{nom}}$ is a vector representing the nominal state of the system, $f(\circ)$ is a function mapping the nominal state vector to the control vector, and ${B}$ represents the attack matrix and nodes from where manipulated measurements are being introduced. The attacker can alter either $B$ or the attack vector, $x_a$, or both to accomplish its objective.

Since the attacker can dynamically alter its attack matrix, the microgrid defender will fail to establish resiliency if it continues to adhere to a single defense strategy. Moreover, continued use of a fixed control policy leaves the microgrid exposed to eavesdroppers who try to snoop on system states and exploit them.

\section{Moving Target Defense} %The issue can be resolved by introducing a dynamically variable MTD control structure that periodically changes the primary control gains .... write that it also has an intrusion detection system that flags anomalous events and triggers the event-triggered secondary gain modification framework to isolate and remove attack injections.
As stated previously, CSAs are often executed in the microgrid environment after careful analysis of system data and control formulations. This allows such attack vectors to learn generic static defenses and bypass them by implementing coordinated manipulations. To counter such attack vectors, it is essential to formulate defenses that can be adaptively altered. This paper uses a hybrid MTD framework to evade, detect, and mitigate CSAs. Evasion is achieved via periodic alteration of primary control gains. Periodic alteration of control gains allows the defender to introduce a level of unpredictability into the system dynamics, making it unfeasible for the attacker to establish an accurate model of the system's operations. Even though evasion is often a good strategy to dodge CSAs, there may be instances where the attacker introduces a strategy that performs successful system manipulation. Hence, we have also included an error computation mechanism that can identify such attacks and trigger the adaptive alteration of secondary-level communication weights to remove manipulations.

At each time step of the MTD execution, choosing the appropriate control gain values is of paramount importance to ensure microgrid stability. The appropriate decision in this context can be reached by strategically analyzing the interactions between the attacker and defender via a game theoretic formulation as depicted below.

\subsection{Game Theoretic Formulation}%%%%%%%% Mention that the game is inherently imperfect information, non-cooperative zero-sum game. However, the digital twin allows the defender to learn the utilities and consequences of its actions in response to the attacker giving it an additional advantage and allowing it to choose the possible action from a given set of defender strategies.
\begin{figure*}
\centering
\centerline{{\includegraphics[width=0.72\linewidth]{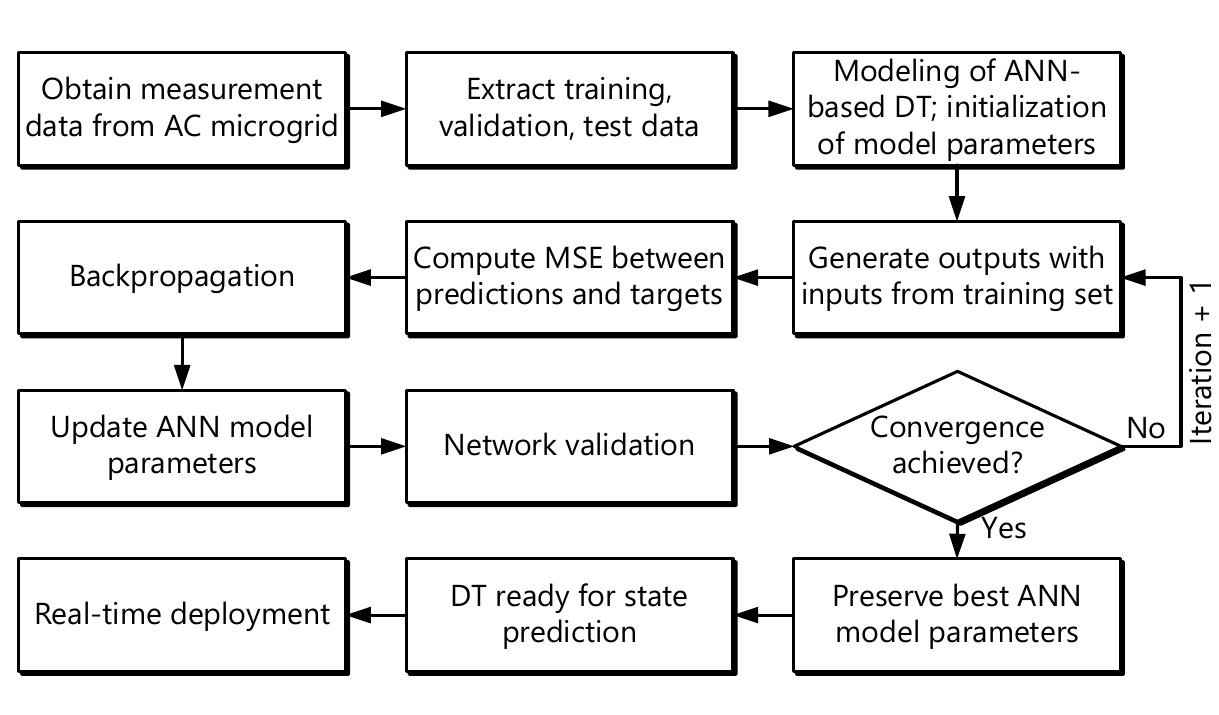}}}
\caption{{Training of the ANN-based DT for state estimation and performance prediction in real-time.}}
\label{fig:train}
\end{figure*}
\begin{figure}
\centering
\centerline{{\includegraphics[width=0.78\linewidth]{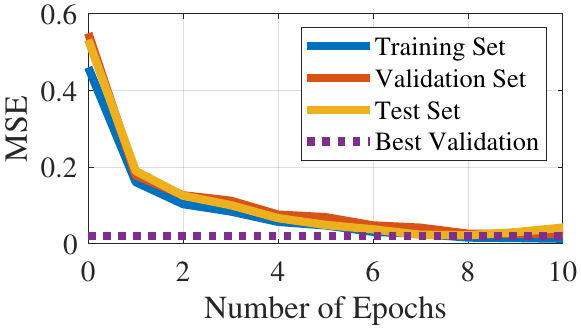}}}
\caption{{MSE values for the ANN-based DT converge to a minimum implying the possibility of higher accuracy during real-time deployment.}}
\label{fig:mse}
\end{figure}
The problem of cybersecurity in the microgrid environment can be best described as a non-cooperative, zero-sum game with the attacker $\mathcal{A_C}$ and the system defender $\mathcal{D_M}$, being the two players. The game is denoted as $\mathcal{G_M} = \{\mathcal{N}_P,\mathcal{S}, \mathcal{U}\}$, where $\mathcal{N}_P:= \{\mathcal{A_C}, \mathcal{D_M}\}$ represents the players, $\mathcal{S}:= \{\mathcal{S_{A_C}}, \mathcal{S_{D_M}}\}$ represents the set of strategies available to each player, and $\mathcal{U}:= \{\mathcal{U_{A_C}}, \mathcal{U_{D_M}}\}$ represents the utilities that can be obtained by the players as a consequence of their actions. The objective of the attacker is to formulate a CSA that can destabilize the system by forcing state parameters to track an uncontrollable trajectory. The attacker also tries to minimize uncertainty in the system as it has to learn microgrid dynamics to formulate the CSA. The attacker's objective $\mathcal{O_{A}}$ can be formulated as:
\begin{equation}
    \mathcal{O_{A}} = min\left(\Xi_F -\sum_{h = 1}^{n_P}v^P_h \right)
\end{equation}
where $\Xi_F$ represents the variation in current state dynamics versus the dynamics in the last time-step and signifies the level of uncertainty in the system, $n_P$ is the number of parameters being evaluated for stability assessment, and $v^P_h$ is the stability degradation metric, which is determined by individual parameters' trajectory deviation (from their nominal path) and the number of oscillations recorded in the last $t$ time slots. $v^P_h$ for each observable parameter is defined as:
\begin{equation}
   v^P_h = z\cdot \frac{p_a}{p_n} + p_r
\end{equation}
where $z$ is the number of observed oscillations, $p_a$ represents average peak-to-peak deviations, $p_n$ is the 
expected parameter magnitude and $p_r$ is the relative error between the current parameter magnitude and expected parameter magnitude \cite{rath2020cyber}. Relative error, $p_r$ is formulated as:
\begin{equation}
    p_r = \frac{|p_c-p_n|}{p_n}
\end{equation}
where $p_c$ represents the current parameter magnitude. This being a zero-sum game, the objective of the defender is to negate the attacker's objective. The defender's objective, $\mathcal{O_{D}}$ can be formulated as:
\begin{equation}
    \mathcal{O_{D}} = max\left(\Xi_F -\sum_{h = 1}^{n_P}v^P_h \right)
\end{equation}
Hence, the cost function $\mathcal{C_D}$ for the microgrid defender can be formulated as:
\begin{equation}
    \mathcal{C_D} = \sum_{h = 1}^{n_P}\left(z\cdot \frac{p_a}{p_n} + p_r\right)_h - \Xi_F
\end{equation}
In microgrid cybersecurity games, the defender may not have the resources to predict whether its actions can achieve the objective in equation (16). To enable the defender to visualize the costs and consequences of its action(s), we design an ANN-enabled virtual twin of the microgrid described in Section III. This visualization can help the defender to compute the utilities corresponding to each strategy available and choose the best response, that is the strategy that will provide the maximum utility. The design and training of the microgrid twin model are explained in the following subsection.
\begin{figure}
\centering
\centerline{{\includegraphics[width=0.92\linewidth]{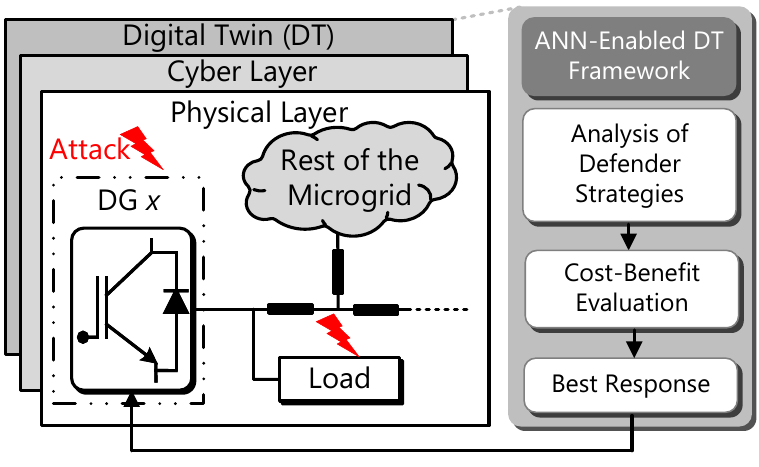}}}
\caption{{Schematic diagram of the proposed DT-aided MTD strategy.}}% Defender decisions are validated by the ANN-enabled DT before real-time execution.
\label{fig:one}
\end{figure}
\subsection{ANN-based DT for Intelligent Decision-Making}
The virtual twin of the microgrid is a multi-input, $L$-layer feed-forward neural network that receives real-time measurements from the system and predicts parameter trajectories in response to possible defender decisions. This mechanism allows the defender to determine possible payoffs for the set of potential strategies and choose the best response to negate the attacker's manipulations. The ANN is trained via supervised learning with labeled historical data generated from an $N$-DG AC microgrid as described in Section III. The inputs in the dataset include two vectors - (i) a vector containing tuples of control gain values, and (ii) a vector containing tuples of state parameter values recorded at the time step just preceding the time slot at which the control gain values were recorded. The output vector in the dataset consists of state parameter values recorded at the same time step as the input vector consisting of control gain values. The format of this dataset is designed to enable the model to learn how to approximate the future values of state parameters when both, a possible set of control gains (potential defender action) and a set of past parameter values are provided. This allows the DT to be used in a multi-purpose manner where it can compute the consequences of defender actions and also predict \texttt{True} values of manipulated signals based on history. The procedure to train the neural network is depicted in Fig. \ref{fig:train}. In the pre-training phase, the ANN is initialized with random model parameters. The training phase consists of several iterations that continue in a loop until convergence is achieved. In each iteration, the model predicts an approximate value of the target output. The weights and biases are updated (via backpropagation) in each iteration until the Mean Squared Error (MSE) between the predicted output vector $\mathbf{X}_{pr}$ and the target vector $\mathbf{X}_{tr}$ converges to a minimum value. For $s_d$ data samples, the MSE is defined as:
\begin{equation}
\text{MSE} = \frac{1}{s_d}\sum_{i=1}^{s_d} (\mathbf{X}_{pr}^{(i)} - \mathbf{X}_{tr}^{(i)})^2
\end{equation}
where both $(\circ)^2$ and $\sum(\circ)$ indicate element-wise operations on the vector.
The MSE values in Fig. \ref{fig:mse} demonstrate the robustness of the DT during the training, testing, and validation phases. The final output vector of the ANN-based DT (at a given time step) can be formulated as a function of the input measurement/control values and the final model parameters:
\begin{equation}
\mathbf{X}_{pr}^{F} = f_A(\psi_w\mathbf{X}_{in}^{F} + \psi_b, L)
\end{equation}
where $\mathbf{X}_{pr}^{F}$ is the final output vector, $\mathbf{X}_{in}^{F}$ represents the input vector, $\psi_w$ and $\psi_b$ represent the final weight and bias vectors respectively, and $f_A(\circ)$ is the activation function applied element-wise. As shown in Fig. \ref{fig:one}, the ANN-enabled DT framework is used to analyze the defender's actions, predict the corresponding payoffs, and determine the best response at all stages where MTD perturbations are to be introduced into the network.

\subsection{Periodic MTD}
The periodic component of the proposed MTD strategy strives to keep the system dynamics unpredictable through the modification of gains associated with the primary control loops, thereby achieving the following objectives:
\begin{itemize}
    \item making it more difficult for the attacker's learning algorithm to establish a clear model of the system's operations,
    \item making the attacker's learned model obsolete, forcing it to learn the system behavior all over again,
    \item increasing the complexity for an attacker to maintain an accurate model of the system's behavior and,
    \item making the attacker incapable of designing CSAs and hiding manipulations by leveraging system knowledge.
\end{itemize}
Another objective that can be achieved by periodically altering the primary control gains is the modification of the system behavior and consequently the thresholds of DG-level bad data detectors. This makes it harder for an attacker to predict the thresholds and bind their injections to avoid detection. The adaptive thresholds can be computed by analyzing the magnitudes of control gains at the $t^{th}$ time step and the expected range of system behaviors as a consequence of these gains.
This paper performs the periodic alteration of primary control gains by adding pseudo-random perturbations to the most recently recorded set of control gain values. Let $K_{PC}$ be a primary control gain. As a consequence of periodic MTD, the magnitude of $K_{PC}$ at the $t^{th}$ time step is formulated as:
\begin{equation}
    K_{PC}(t) = K_{PC}(t-1) + K_{PRN}(t)
\end{equation}
where $K_{PRN}$ is a pseudorandom value that varies with time. $K_{PRN}$ is formulated as:
\begin{equation}
    K_{PRN}(t) = g(t, \zeta_d)\cdot\zeta_e
\end{equation}
where $g(\circ)$ is the pseudo-random number generation function, $\zeta_d$ is a secret value set by the defender that initializes $g(\circ)$, and $\zeta_e$ is a constant scaling parameter that determines the maximum magnitude of the alteration. After the computation of $K_{PC}(t)$, it is introduced into the microgrid-digital twin environment and its impact on the system dynamics is observed. If the observed state trajectory is deemed acceptable, the perturbed control gain is introduced into the actual microgrid setup.

Even though the periodic alteration introduces an element of uncertainty into the control framework to enhance the cost and diminish the success probability of attacking the microgrid, the rapidly evolving world of cyber-infiltration has made it impractical to assume complete evasion against all attack types. Hence, the MTD framework also consists of an attack detection and event-triggered mitigation mechanism, that isolates and removes embedded manipulations. The principle of operation for the anomaly detection and mitigation framework is explained in the following subsection.

\subsection{Anomaly Detection and Event-Triggered MTD}
In the normal scenario, the distributed microgrid setup achieves synchrony among nodal measurements via cooperative control \cite{rath2020cyber}. The inability of the attacker to hide its manipulations (by leveraging system knowledge) makes it easier for the defender to exploit this physical property and create a localized rule-based detection metric to detect anomalies at the DG level. On the receipt of an incoming measurement signal (from a neighboring DG), the local controller at the receiving node calculates the error margin, $e$ between the signals from other transmitting nodes and its own local measurements to determine if the signal is breached. Error margin $e$, which also acts as the trigger signal for the event-triggered MTD is formulated as:
\begin{equation}
    e = a_{xy}(x_y-x_x)^2 + a_{xz}(x_z-x_x)^2 + ...
\end{equation}
where $a_{xz}$ is the weight assigned to the signal received from the $z^{th}$ node, and $x_z$ denotes the local measurements corresponding to the $z^{th}$ node.

Post-computation of $e$, the controller compares it against a locally generated threshold margin, $e_{th}$. If $|e|>e_{th}$, an anomalous flag is raised that activates the event-triggered component of the MTD framework.
In a practical, real-world setup $e_{th}$ can be formulated as an infinitesimally small value that is established by the defender by analyzing historical microgrid data to determine the maximum range of nominal asynchrony caused by natural phenomena (e.g., noise).
After the receipt of the trigger signal, the event-triggered MTD framework responds by perturbing the communication weight(s) ($a_{x\circ}$ in equations (10) and (11)) in an attempt to reduce the weight assigned to the \texttt{False} measurements to zero, thereby completely eliminating their impact on the microgrid control framework. After attack removal, the defender uses the ANN-based DT to reconstruct \texttt{True} values of the manipulated signals and replaces them. The proposed strategy achieves $N$-resiliency by reconstructing \texttt{True} values of the manipulated signals using the ANN-based DT and replacing the magnitude of the \texttt{False} signal in equations (10) and (11) to continue nominal operations.

\section{Performance Evaluation and Results}
\begin{figure}
    \centering
    \includegraphics[width=0.45\linewidth,clip,trim={6 6 6 121}]{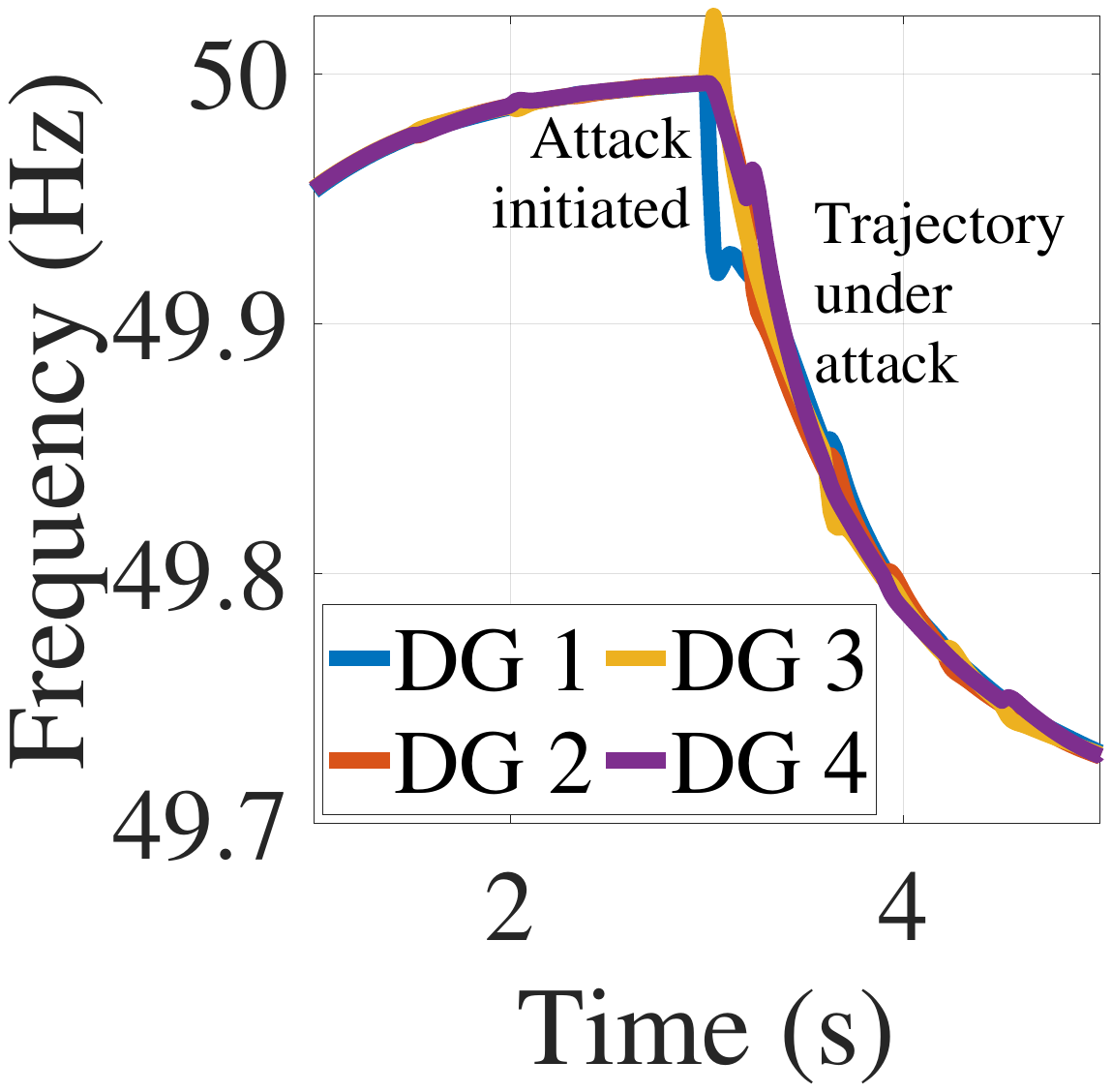}
    \includegraphics[width=0.45\linewidth,clip,trim={6 6 6 121}]{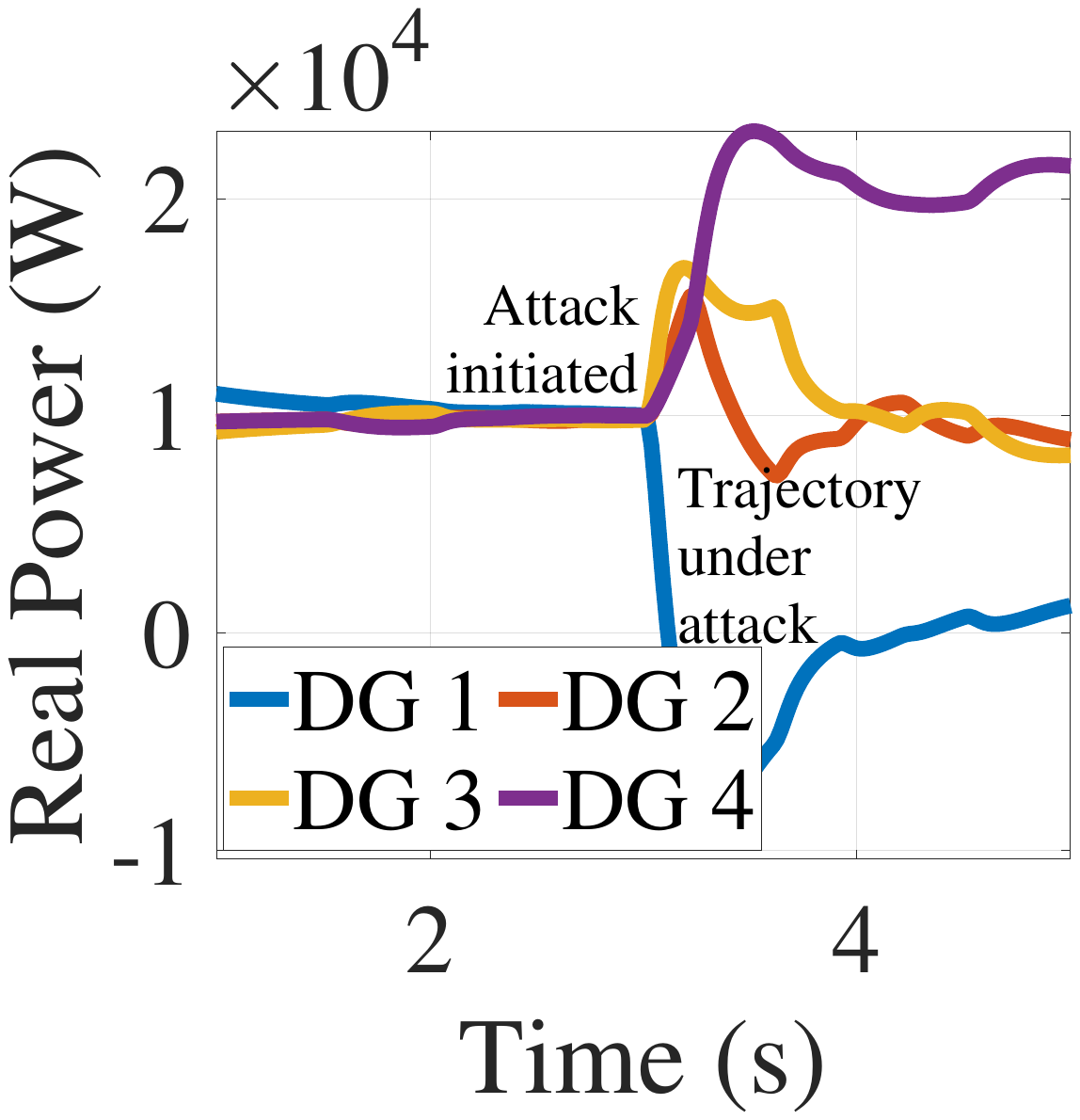}\\[-4ex]
    \includegraphics[width=0.45\linewidth,clip,trim={6 6 6 121}]{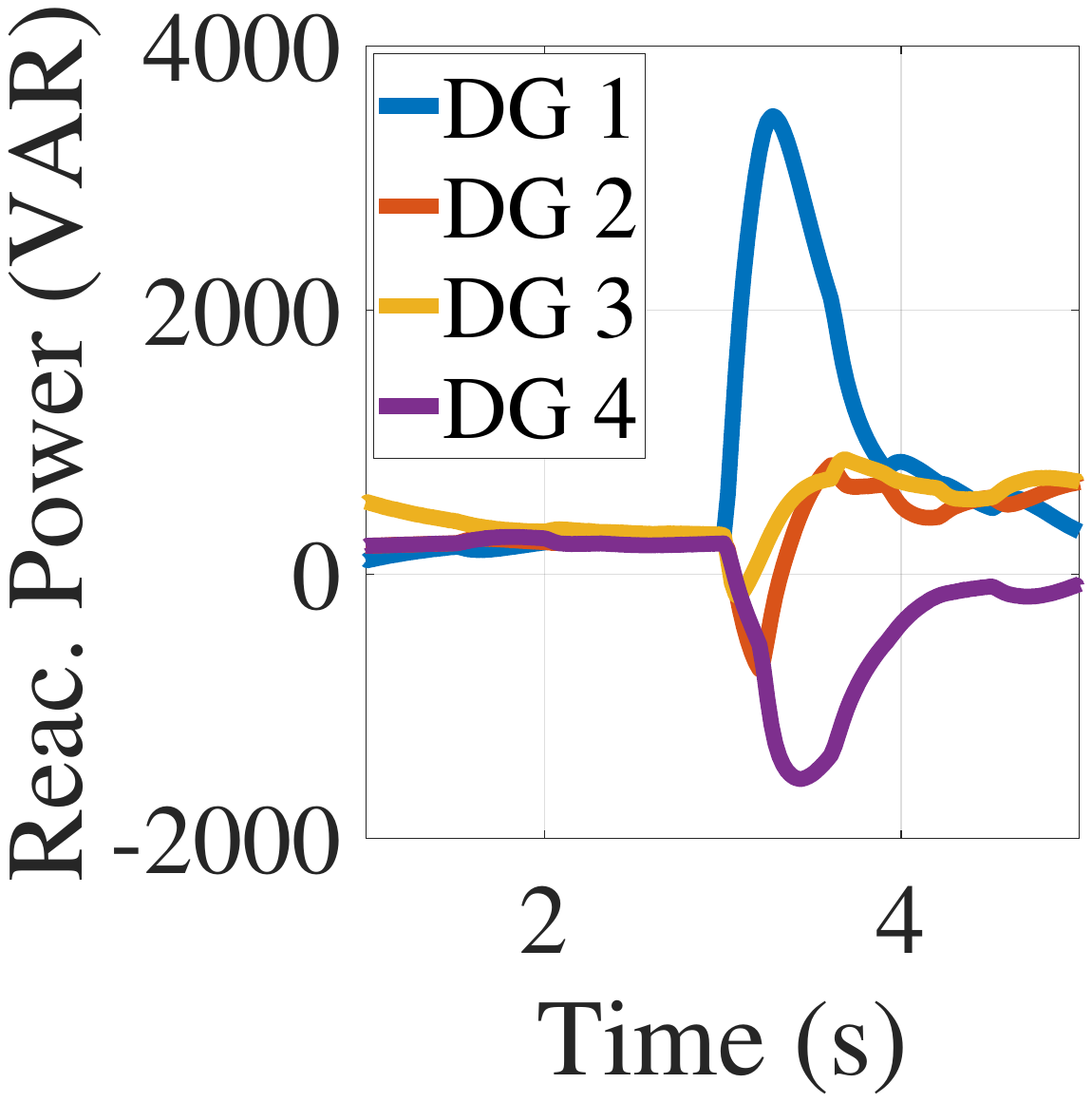}
    \includegraphics[width=0.45\linewidth,clip,trim={6 6 6 141}]{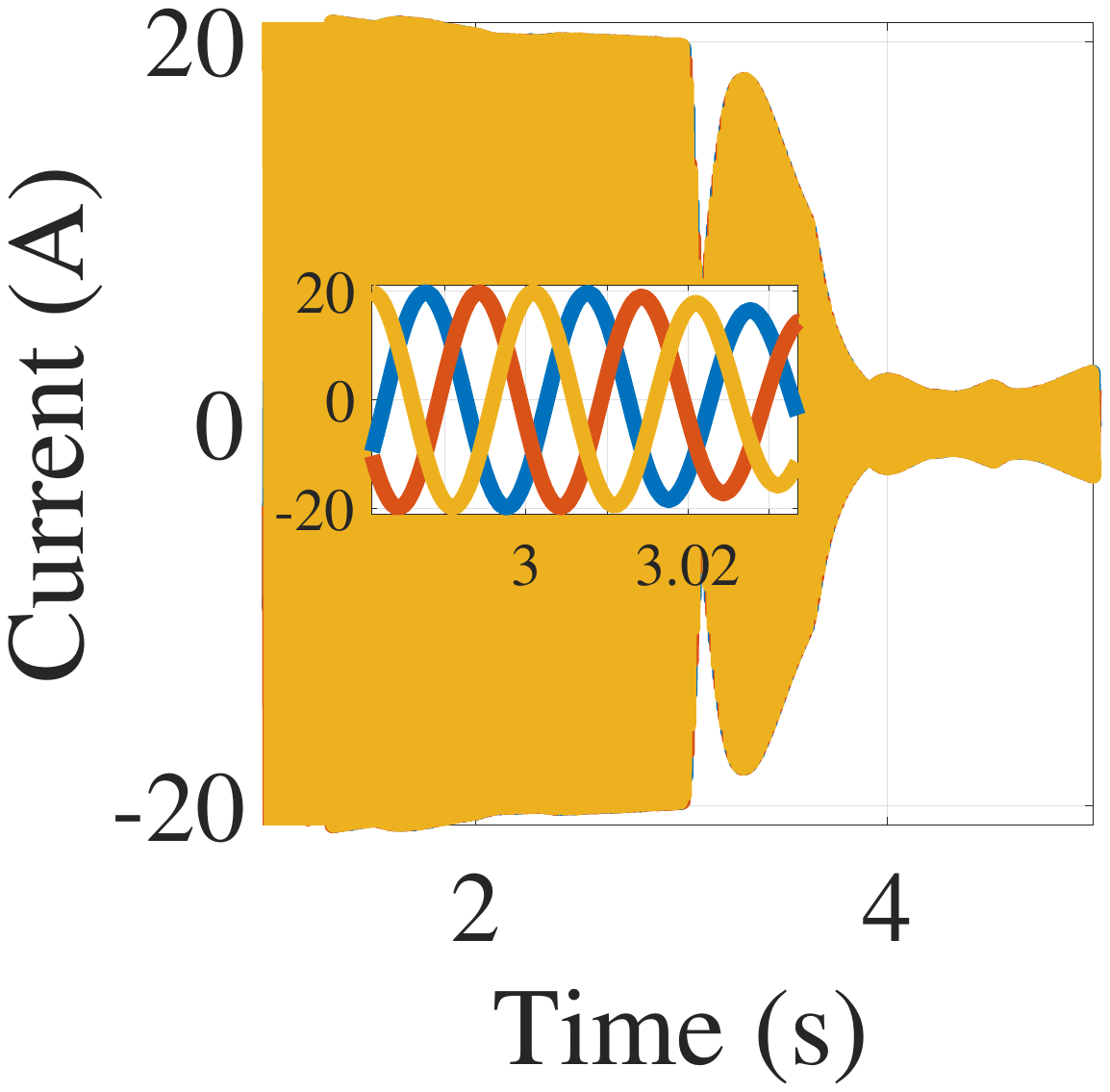}\\[-4ex]
    \includegraphics[width=0.45\linewidth,clip,trim={6 6 6 121}]{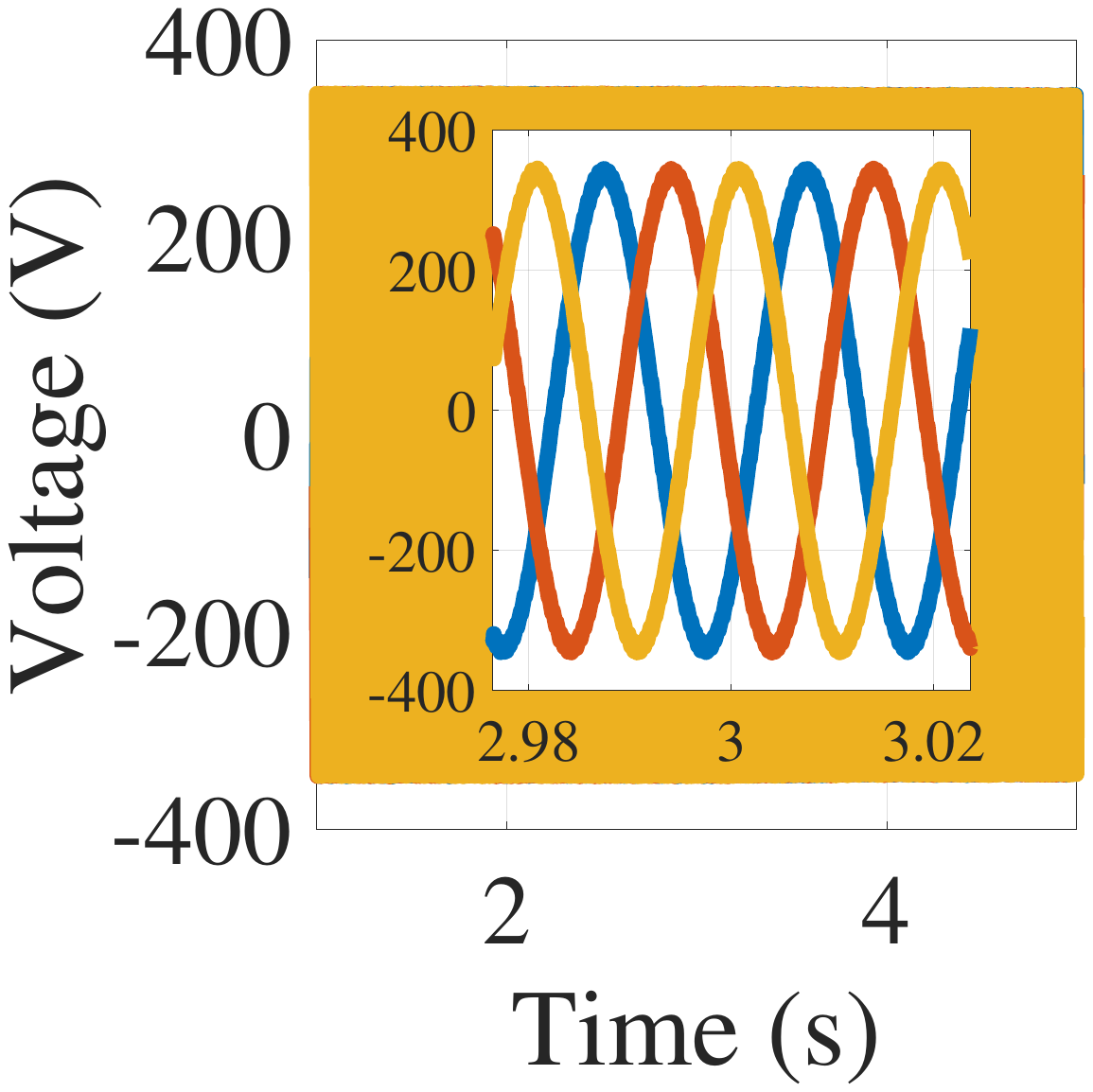}
    \includegraphics[width=0.45\linewidth,clip,trim={6 6 6 141}]{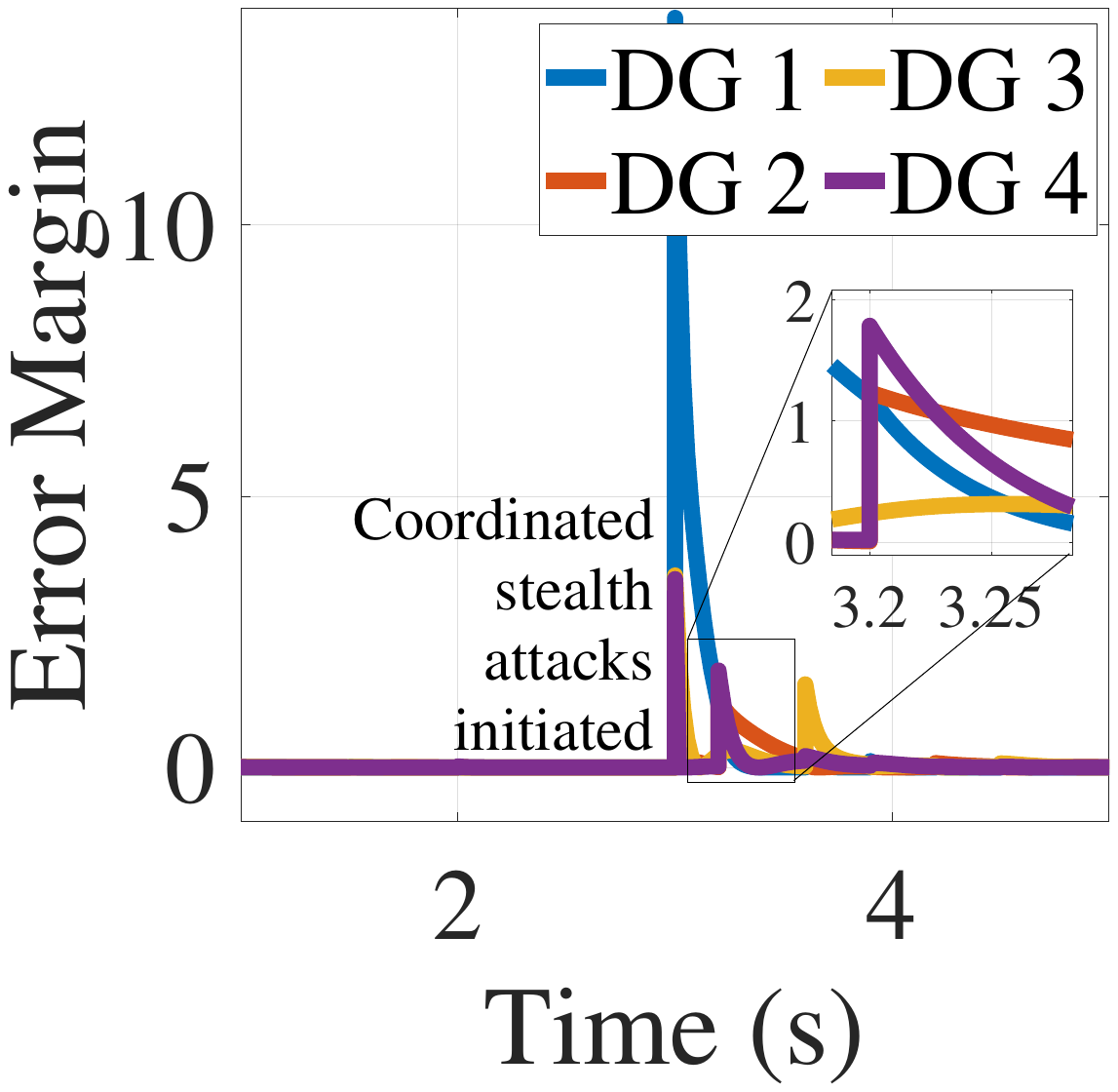}\\[-4ex]
    \caption{Performance of a generic microgrid (not equipped with MTD) in the presence of a CSA.}
    \label{fig:result1}
\end{figure}

\begin{figure}
    \centering
    \includegraphics[width=0.45\linewidth,clip,trim={6 6 6 121}]{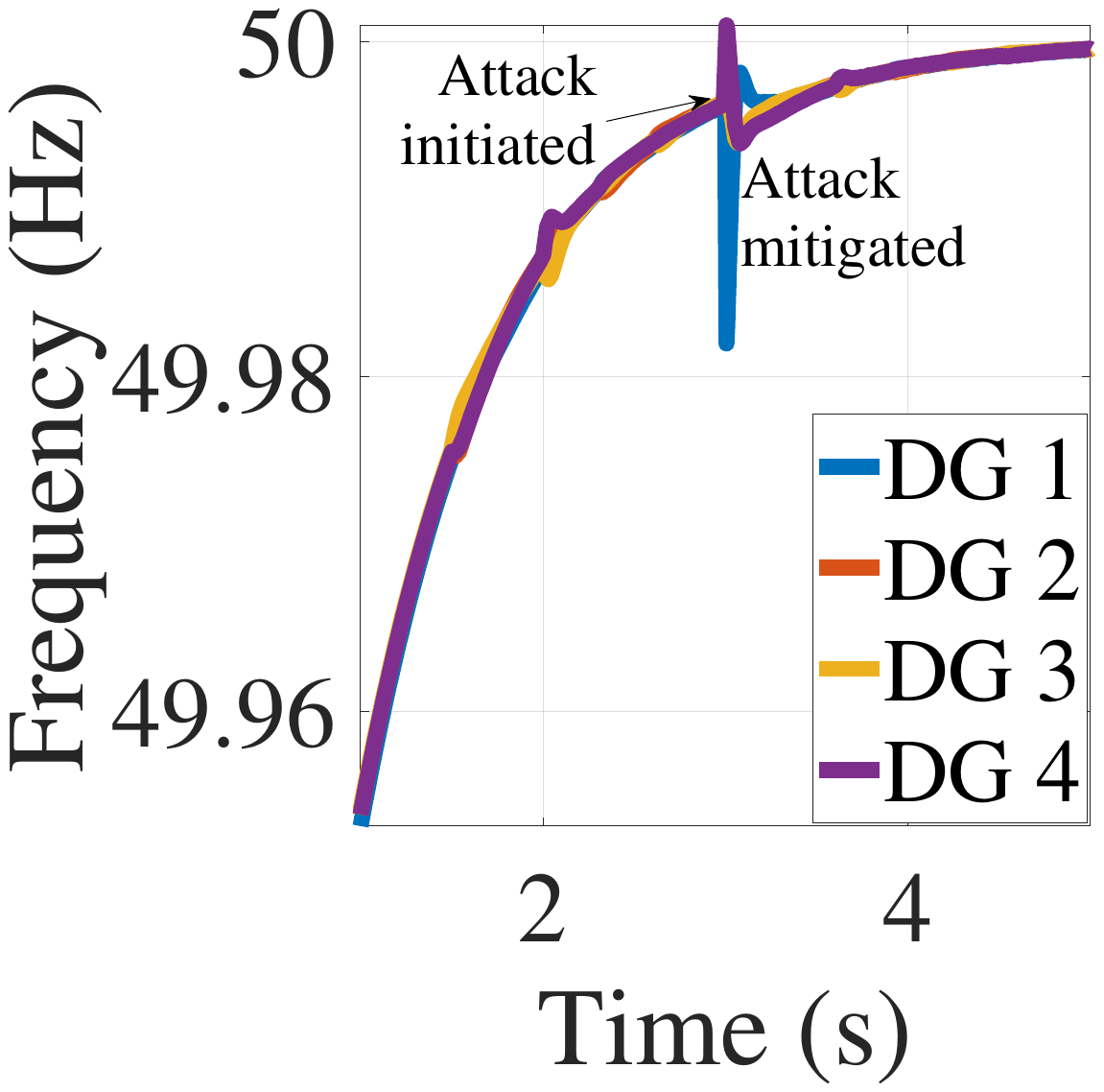}
    \includegraphics[width=0.45\linewidth,clip,trim={6 6 6 121}]{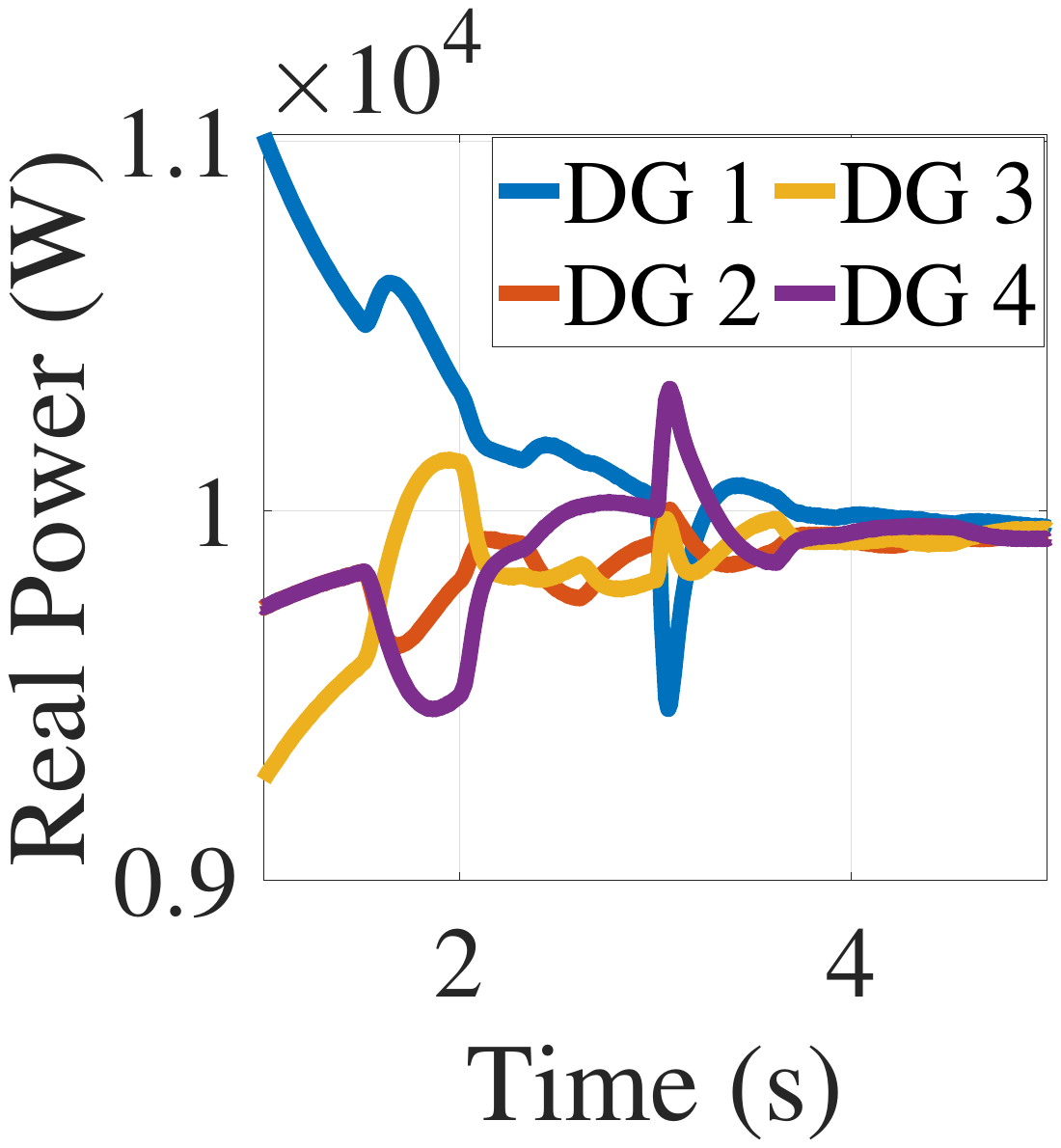}\\[-4ex]
    \includegraphics[width=0.45\linewidth,clip,trim={6 6 6 121}]{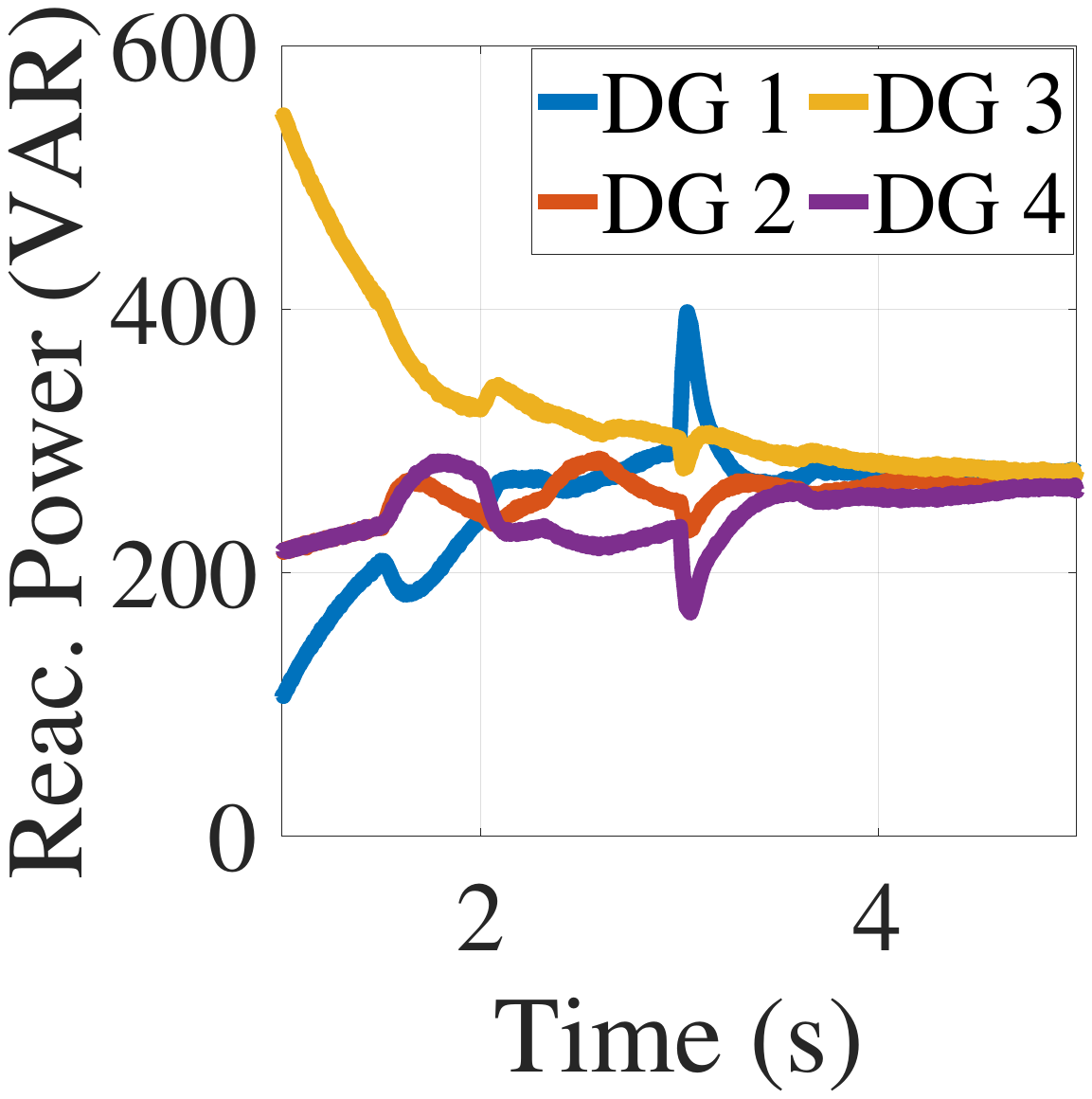}
    \includegraphics[width=0.45\linewidth,clip,trim={6 6 6 141}]{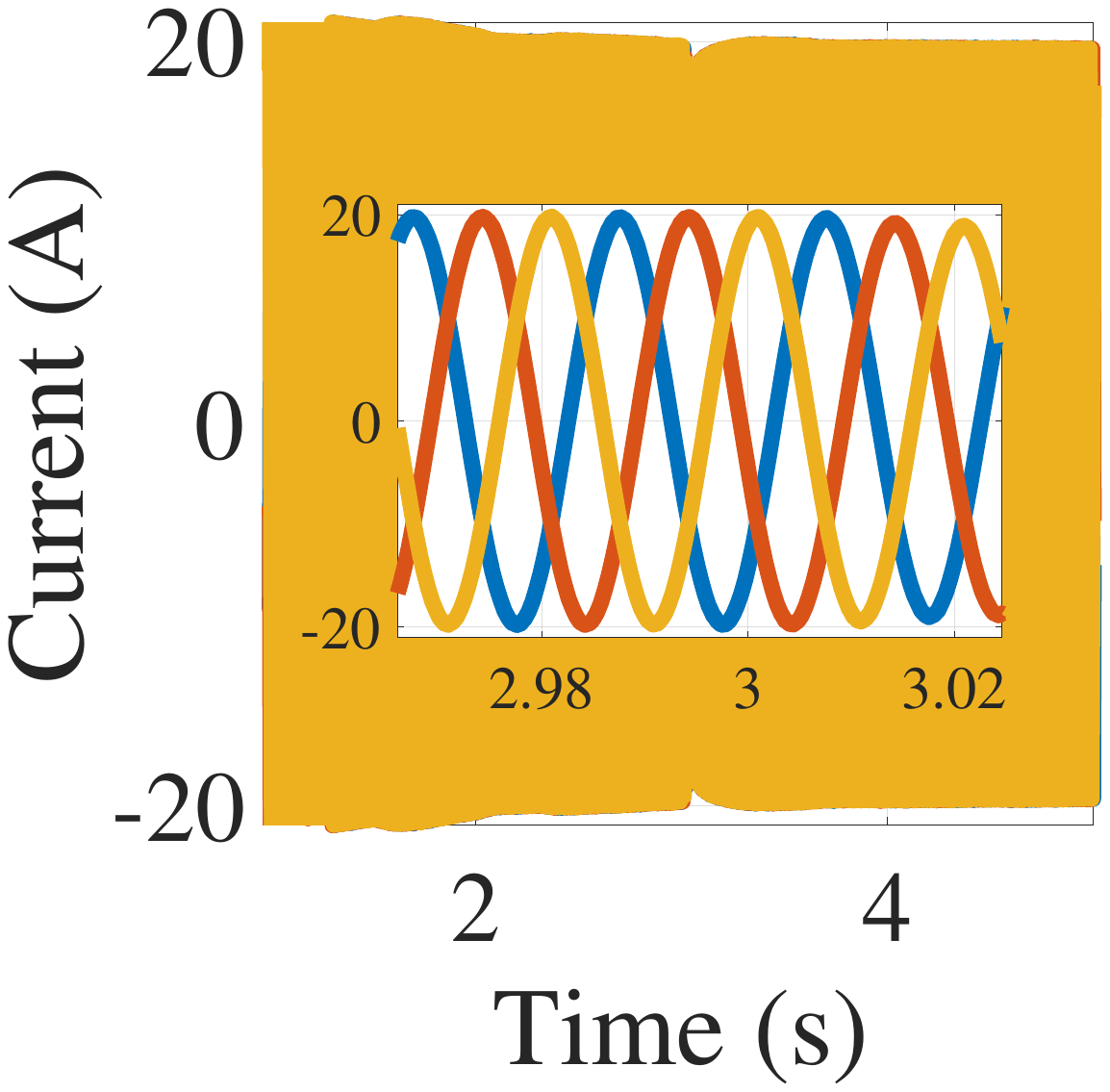}\\[-4ex]
    \includegraphics[width=0.45\linewidth,clip,trim={6 6 6 121}]{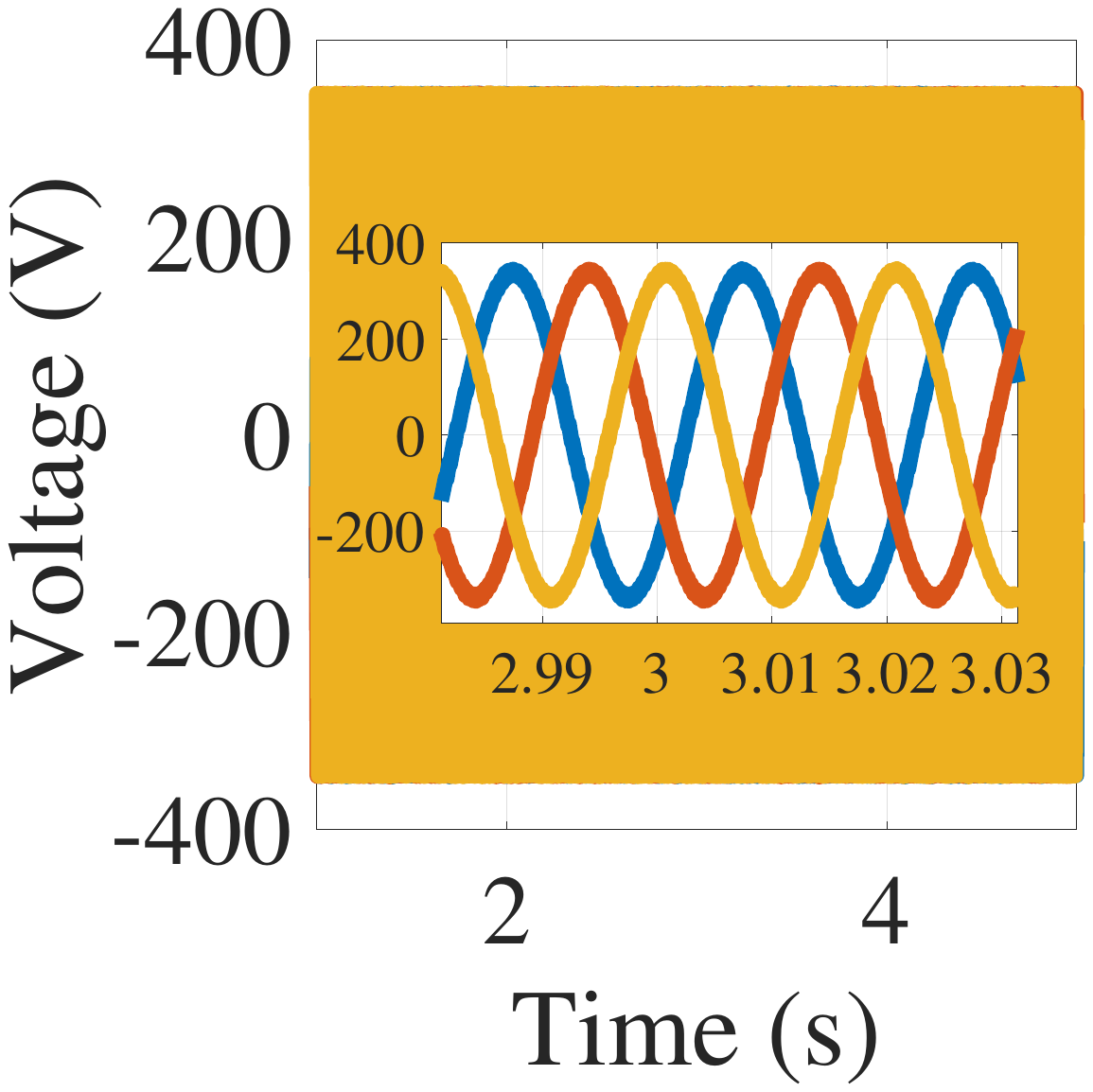}
    \includegraphics[width=0.45\linewidth,clip,trim={6 6 6 141}]{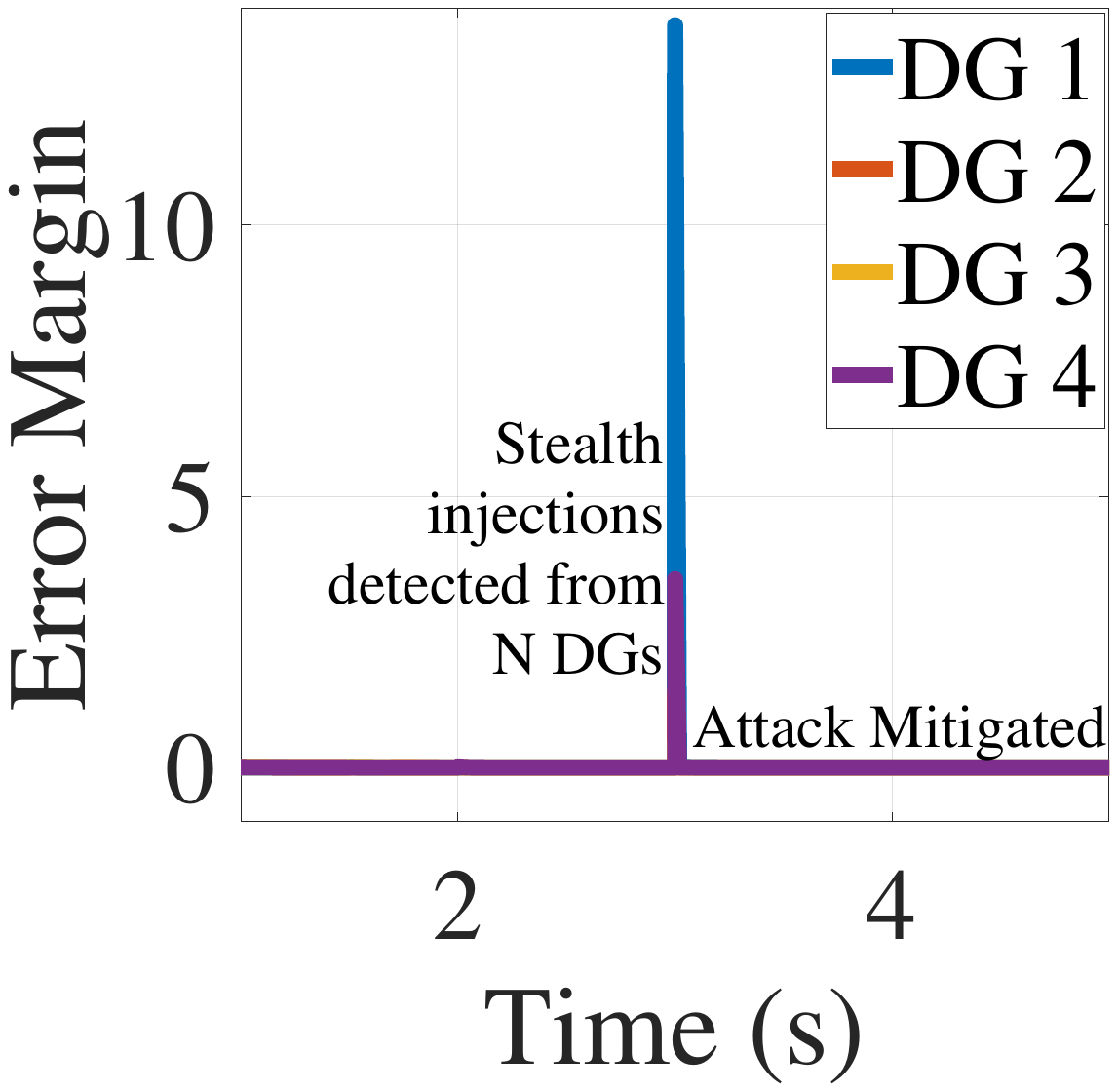}\\[-4ex]
    \caption{Performance of the hybrid MTD mechanism in the presence of a CSA introducing manipulations from all four DGs in the network ($N$-resiliency).}
    \label{fig:result2}
\end{figure}
The test system used for evaluating the efficacy of the proposed MTD strategy is an autonomous, 4-DG AC microgrid with a hierarchical control structure as depicted in Section III. The testbed is implemented in the MATLAB R2020a environment. The model parameters for this system can be obtained from \cite{rath2020cyber}. A DT for the modeled microgrid which is essentially a feed-forward, densely connected ANN with four layers (including the input and output layers) is built in the Python environment. The ANN is trained in a supervised manner with historical microgrid data which is generated from the test system depicted above. The labeled training dataset is composed of two input vectors and one output vector, which include several sets of microgrid state parameters and local, DG-level control gain values. The input vectors are concatenated before being passed through the hidden layers, where each unit utilizes a Rectified Linear Unit (ReLU) activation function. The model is compiled using the Adam optimizer and utilizes MSE as the loss function. To analyze the robustness of the proposed strategy, we consider two case studies - (i) exposure of the generic microgrid (which is not configured to use the DT-aided MTD strategy) to a CSA targeting all four nodes in the network, and (ii) exposure of a microgrid fortified with the proposed DT-enabled MTD to the CSA used in the first case study. More details about the case studies are provided below.

\subsection{Case Study I}
For this case study, the cyber-physical microgrid structure as depicted above is not configured to access the DT. The system is exposed to a CSA that initiates malicious manipulations from all four DGs in the network simultaneously (time of initiation, $t = 3$ s). Fig. \ref{fig:result1} showcases the performance of the system before and after the attack initiation. The attack vector successfully disrupts the microgrid states, causing an imbalance in real and reactive power sharing among the DGs. The attack vector also creates a deviation in the trajectory of local frequencies measured at the DGs. The magnitude of current as measured at DG 1 is gradually reduced as a consequence of the elongated exposure to the CSA. If the attack persists for a longer time period, it may be able to enhance the imbalance in the microgrid states to cause a complete system failure.

\subsection{Case Study II}
\begin{figure}
    \centering
    \includegraphics[width=0.45\linewidth,clip,trim={6 6 6 115}]{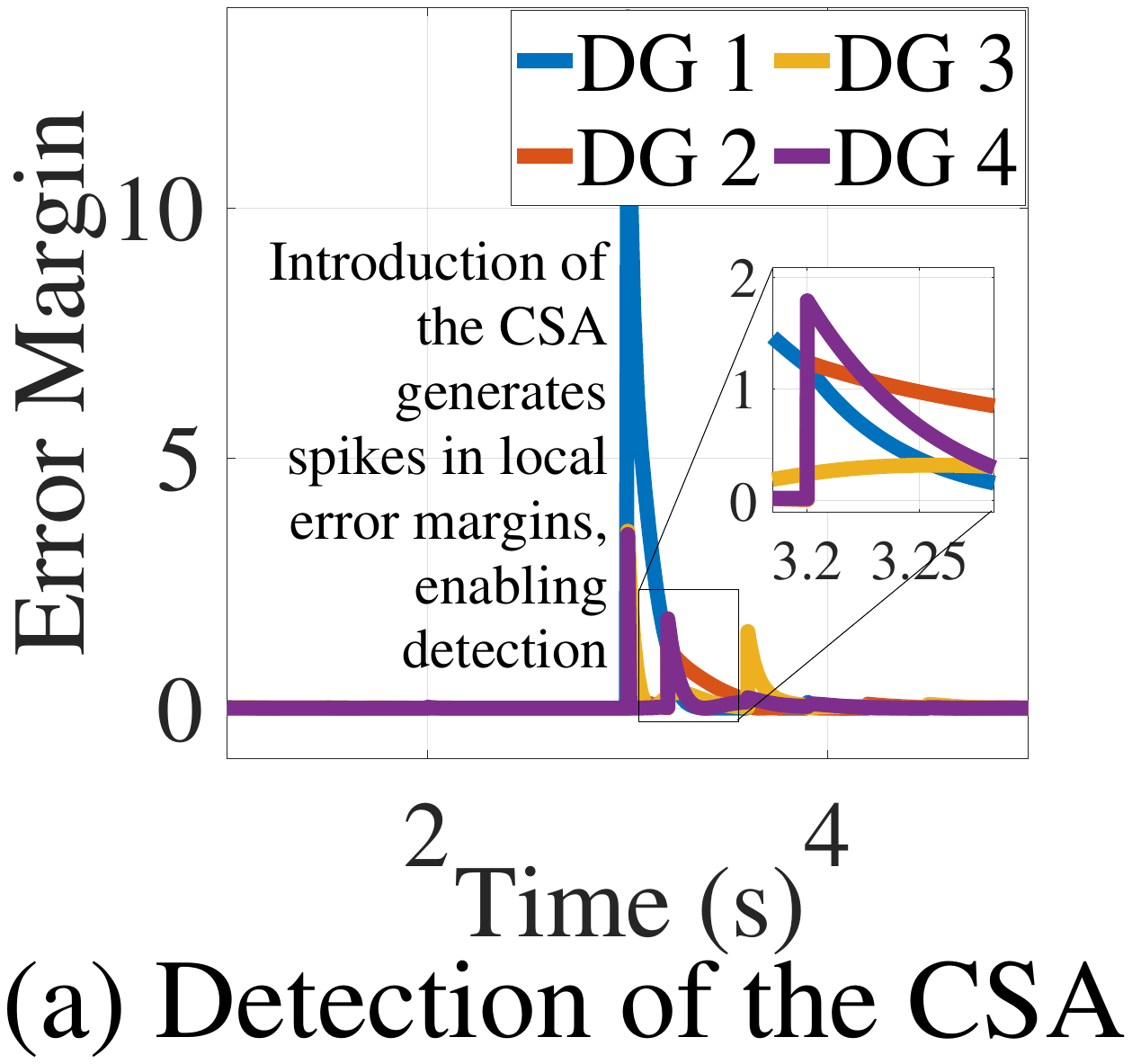}
    \includegraphics[width=0.45\linewidth,clip,trim={6 6 6 107}]{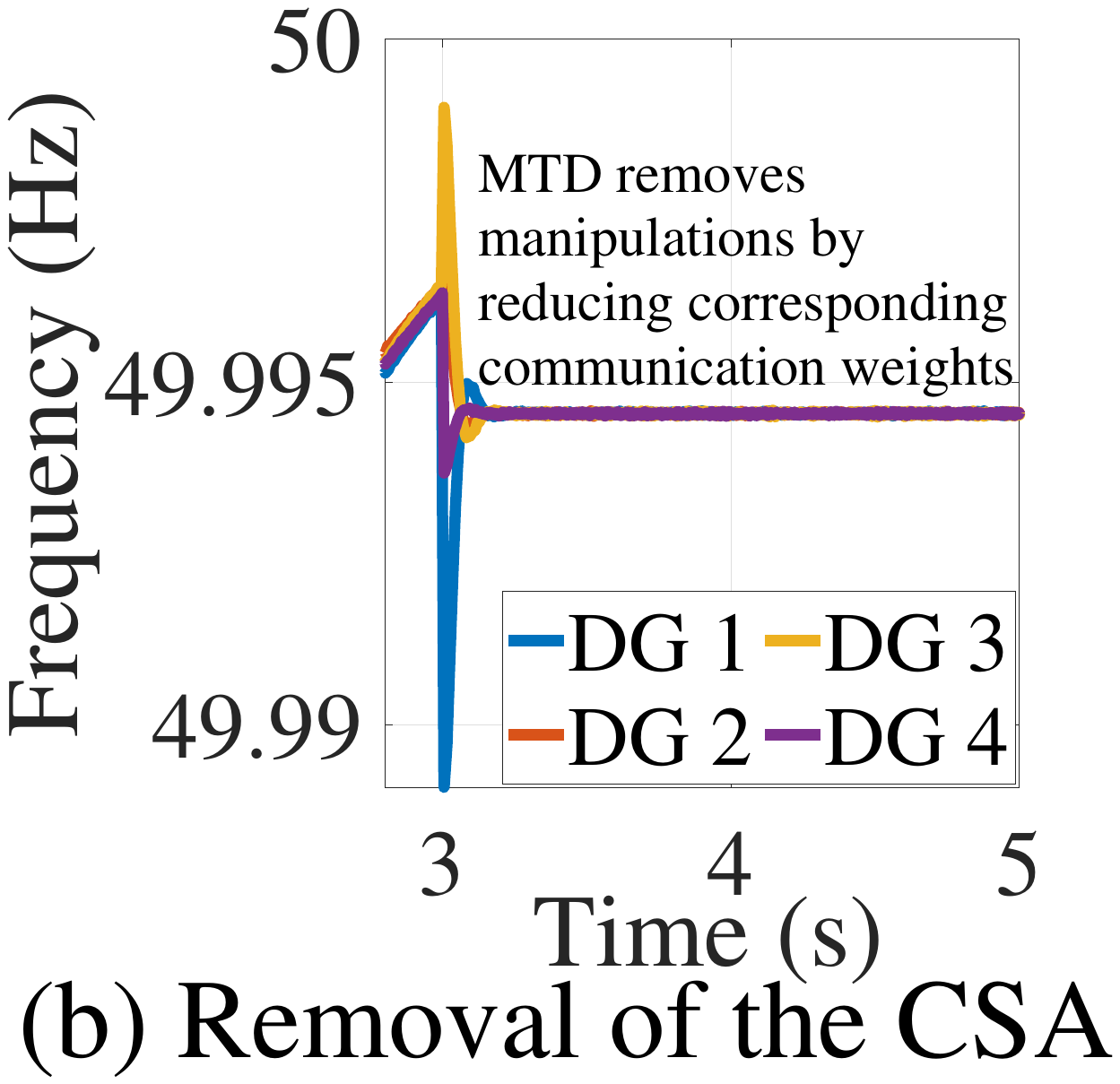}\\[-4ex]
    \includegraphics[width=0.45\linewidth,clip,trim={6 6 6 109}]{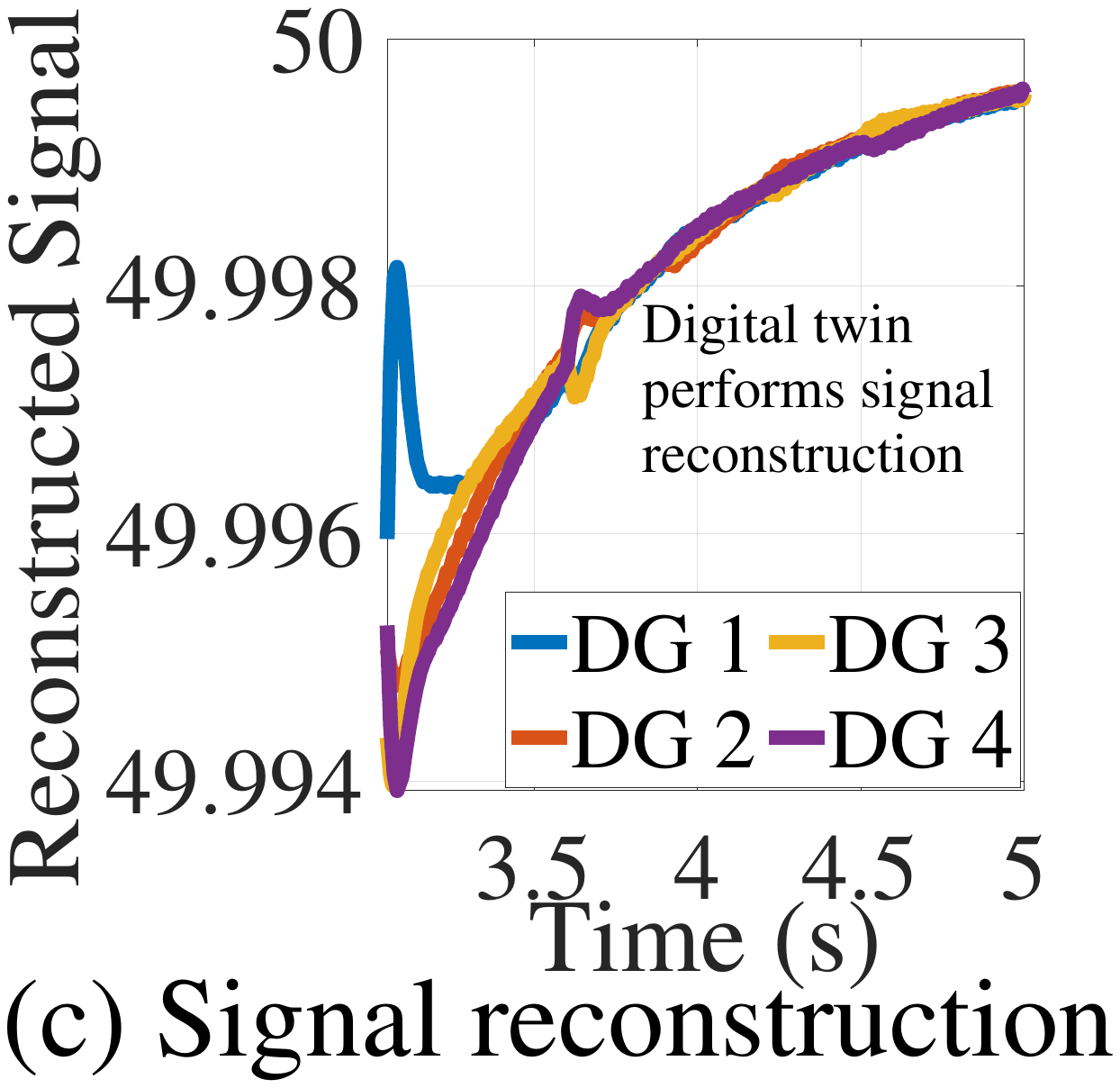}
    \includegraphics[width=0.45\linewidth,clip,trim={6 6 6 107}]{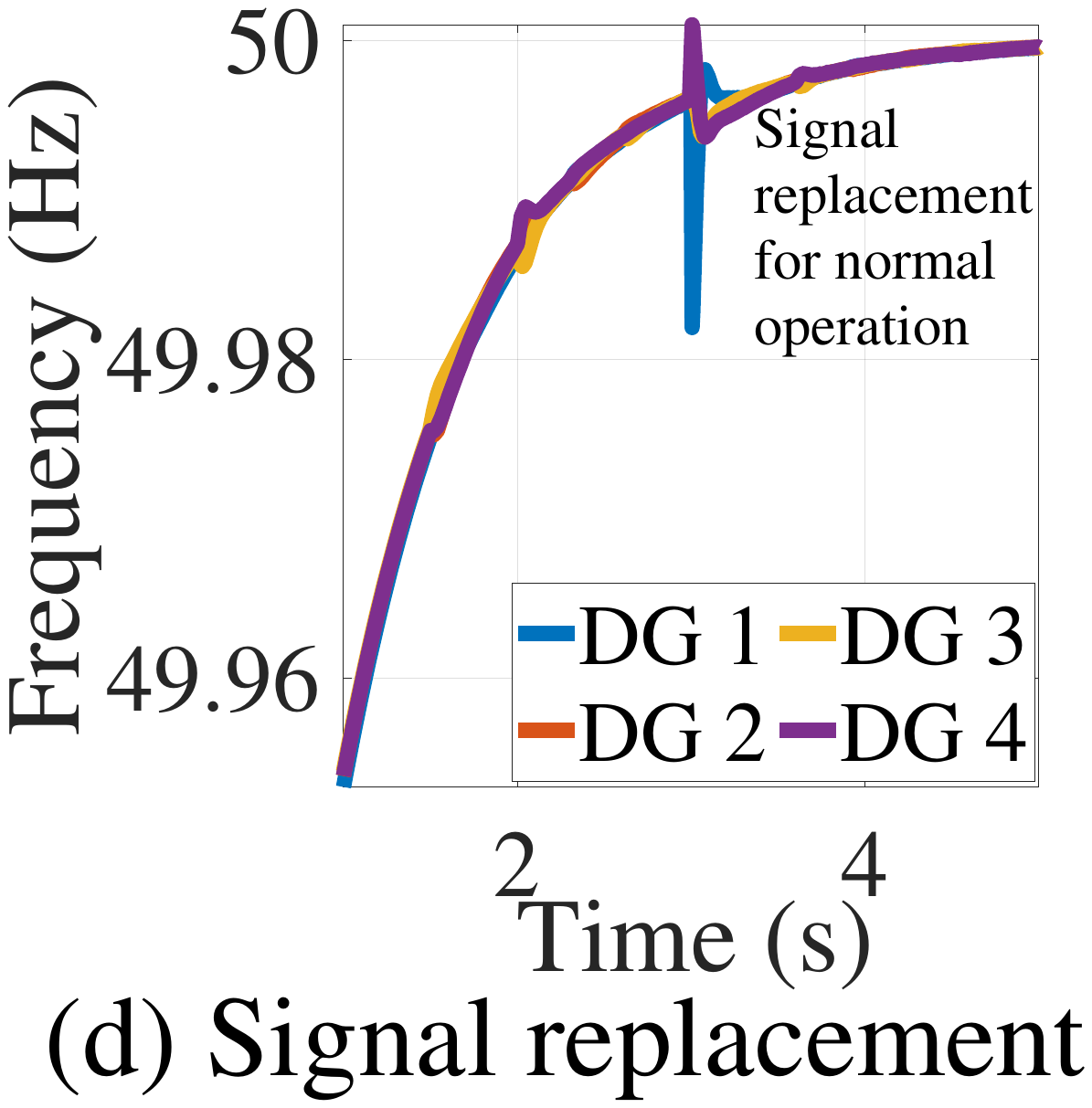}\\[-4ex]
   \caption{Schematic diagram depicting the steps involved in the proposed mitigation framework.}
   \label{fig:result3}
\end{figure}

For this case study, the microgrid system as depicted in this section is configured to access the DT and perform periodic and event-triggered MTD as per the strategy depicted in Section IV. The CSA used in Case Study I is injected into the system to initiate manipulations from all four DGs simultaneously (at $t = 3$ s). As depicted in Fig. \ref{fig:result2}, the activation of the CSA creates a very high spike for all the locally computed $e$ values, triggering an instantaneous activation of the event-triggered MTD framework. Once triggered, the MTD framework utilizes the DT to determine a possible combination of communication weights (in the secondary layer) that can remove all the manipulations introduced by the attacker and limit attack propagation. After removing the attack, the system uses the ANN-based DT to reconstruct the lost signals and feeds them to the affected controllers enabling them to function normally without disrupting the nominal state of the microgrid.
A detailed explanation of the proposed mitigation framework is depicted in Fig. \ref{fig:result3}. As mentioned before, the detection of the CSA triggers a series of steps that ultimately result in attack removal. The system recovers the removed signals by reconstructing them using the ANN-enabled DT.

\section{Conclusion}
CSAs can affect the stability of cyber-physical microgrids, forcing their states to collapse into an uncontrollable trajectory. Such attack vectors often tend to hide from the defenders' radar by creating low-level, bounded manipulations from several locations at the same time. This paper used the concept of game theory to model and analyze the interactions between a CSA and a microgrid defender as a zero-sum, non-cooperative game. Additionally, it proposed a hybrid MTD strategy (consisting of periodic and event-triggered components) to evade, detect and mitigate CSAs in AC microgrids. The periodic layer of the proposed strategy was utilized to reduce the success probability of attacks through the perturbation of primary control gains. Periodic MTD also made it difficult for attackers to hide manipulations allowing defenders to detect them via simplified rule-based detection criteria. Post-detection, the event-triggered component of the proposed strategy was used to alter communication weights assigned to malicious injections (in the secondary control layer) to neutralize their impact. Acknowledging that miscalculated reconfigurations in the microgrid control plane can destabilize the system, all MTD decisions were validated using an ANN-based DT before the execution of the best response in the actual system.
The ANN-based DT was also used to reconstruct the manipulated signal(s) to protect system stability and achieve $N$-resiliency. Several simulation results were provided validating the effectiveness of the proposed strategy against CSAs. Future extension of this paper will attempt to study the impact of adversarial data-poisoning attacks on ANN-based DT models and formulate a strategy to perform supervised training of such models in low-trust, adversarial settings.

\section*{Acknowledgment}

This research is supported by the National Science Foundation (NSF) Award No. 2019164. % Readers may contact srath@nevada.unr.edu for requesting access to the codes and dataset used in this paper.

%\balance
\bibliographystyle{IEEEtran}
\bibliography{biblio}

\end{document}